# The Design and Testing of the Address in Real Time Data Driver Card for the Micromegas Detector of the ATLAS New Small Wheel Upgrade

L. Yao, H. Chen, K. Chen, S. Tang, and V. Polychronakos

*Abstract*— The Address in Real Time Data Driver Card (ADDC) is designed to transmit the trigger data in the Micromegas detector of the ATLAS New Small Wheel (NSW) upgrade. The ART signals are generated by the front end ASIC, named VMM chip, to indicate the address of the first above-threshold event. A custom ASIC (ART ASIC) is designed to receive the ART signals from the VMM chip and do the hit-selection processing. Processed data from ART ASIC will be transmitted out of the NSW detector through GBTx serializer, VTTx optical transmitter module and fiber optical links.

The ART signal is critical for the ATLAS experiment trigger selection thus the functionality and stability of the ADDC is very important. To ensure extensive testing of the ADDC, an FMC based testing platform and special firmware/software are developed. This test platform works with the commercial Xilinx VC707 FPGA develop kit, even without the other electronics of the NSW it can test all the functionality of the ADDC and also long term stability. This paper will introduce the design, testing procedure and results from the ADDC and the FMC testing platform.

## I. INTRODUCTION

THE New Small Wheel (NSW) is part of the ATLAS Phase I upgrade project. This NSW is planned to replace the present Muon Small Wheel and provide the capability to meet the increased luminosity and reduce the fake triggers [1][2]. In the NSW upgrade, two different types of detectors are used: the small-strip Thin Gap Chambers (sTGC) and the MicroMegas (MM). The signals from both detectors will contribute in the trigger decision and by crosscheck most of the fake triggers can be eliminated [3].

In the MicroMegas detector the Address in Real Time (ART) signal is used to generate the trigger decision [4]. This signal is generated by the 64-channel front end ASIC, VMM chip, when a detector channel is hit and the output pulse goes above a given threshold [5]. In every bunch crossing the VMM chip will only provide an encoded 6-bit address of the first hit channel. This encoded 6-bit address is the Address in Real Time, namely ART signal.

The block diagram of the NSW electronics is shown in Figure 1 [6]. The sTGC and MM detector will have its own electronics for data transmission. In MM detector electronics, the Level-1 data (time, charge and strip address corresponding to a single hit) will be transferred through the Level-1 Data Driver Card (L1DDC) to a network interface called Front End LInk eXchange (FELIX) [7][8]. The ART data driver card (ADDC) is in the trigger path to transfer the ART signals from the front end boards to the Trigger processor for the trigger decision. To reduce the pressure on the Trigger processor, the ADDC also need to perform preliminary hit-selection on board. The trigger decision from the NSW is part of the ATLAS global trigger system and the latency needs to be controlled within ~1us [9]. Several simulation have been performed and the preliminary conclusion is that about 500 ns is required for the trigger processor to make the trigger decision, leaving about 500 ns for the ART data transmission and processing by the ADDC [10].

All the NSW electronics will be synchronized to the global 40.079 MHz bunch crossing clock. This clock will be generated by the TTC module and distributed through the FELIX. The FELIX will also distribute the slow control commands. To reduce the complexity, the ADDC does not receive the clock and command signals directly from FELIX, there will be a cable connection between L1DDC and ADDC and these signals will be transferred to ADDC through this link.

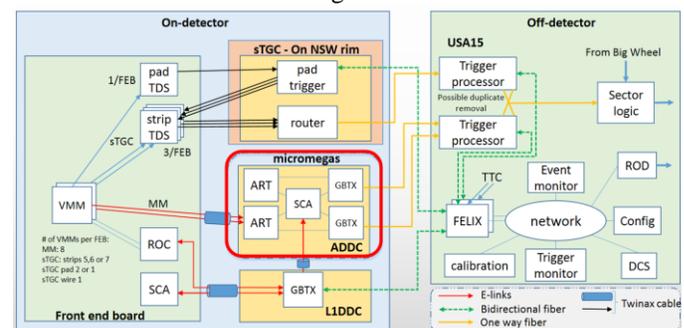

Fig. 1. The block diagram of the NSW electronics. ADDC is on the MicroMegas Detector trigger path to transmit the trigger primitive, ART signal, from front end to the Trigger processor.

To reduce the ART signal attenuation, the ADDC will be placed close to the MicroMegas Front End (MMFE) boards on the detector rim. The estimated radiation dose is up to 1700 Gy

---

(inner radius) and the magnetic field is up to 0.4 T. To guarantee the performance and stability under such environment, several custom ASICs are adopted. The ART signals will be processed by a custom ASIC, the ART ASIC and each one can receive up to 32 ART signals from 32 VMM chips on 4 different MMFE boards. The processed data will be collected by the GBTx serializer ASIC and the unidirectional Versatile Twin Transmitter (VTTx) module will transmit them to the trigger processor through optical links. To take advantage of the dual optical transmitter links on one VTTx module, each ADDC board is designed to handle 64 ART signals with 2 ART ASICs and 2 GBTx chips. Another ASIC of the GBT chipset, the SCA slow control chip, will be used for the configuration and control of the other chips on ADDC. In the NSW upgrade plan, a total of 512 ADDC boards will be used to connect the 4096 front end boards.

This paper will introduce the design of the ADDC and the history of its prototypes. To accelerate the testing of the ADDC, a test platform is designed based on the VC707 commercial FPGA develop kit. The design of this test platform is introduced and the results, including the functionality test results and the latency performance, will also be presented.

## II. ADDC DESIGN AND PROTOTYPES

The ART signal generated by the VMM front-end ASIC consists of a flag pulse, followed by 6 bits which indicate the address of the strip that received a hit. The internal circuitry of the VMM selects always the first channel that was hit, when more channels receive a signal nearly at the same time. The ART circuit is triggered either by the threshold crossing of the strip signal or by the peak detection circuit. For minimum latency, the threshold crossing will be used in the experiment. The flag is high for two falling edges of the 160 MHz clock and kept low until the next rising edge of the clock. The 6 address bits are then serialized on each edge of the clock. Following the ART address, the ART circuit is internally reset for approximatively 10ns [11].

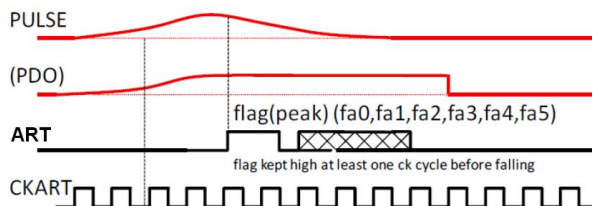

Fig. 2. The ART signal will be generated by the front end ASIC, VMM chip, after a signal pulse reaches its channel. The ART signal records only the first hit channel number.

The dimensions of the ADDC are limited to 200 mm x 60 mm and the overall height, including component and board thickness, should be less than 15 mm. Each ADDC board should be able to handle 64 channels of ART stream input and process the hit selection function. This function includes:
• Deserialize the ART stream and phase-align the hits to the BC clock.
• Identify the strip addresses of up to 8 of hits by means of cascaded priority encoders.
• Append the 5-bit geographical VMM ASIC address to the strip address of each hit.
• Send the ART addresses and the 12-bit BCID to the trigger processor, the data output must be in GBT widebus format.

The board also should be able to be synchronized with the LHC BC clock (40.079 MHz). It should be able to receive remote configuration and monitoring from FELIX. The input voltage for board power supply is 12V.

The final ADDC structure is shown below in figure 3. As introduced above, each ADDC can receive 64 channels of ART inputs from 8 miniSAS connectors. Each miniSAS connector connects with a MMFE front end board (with 8 VMM chips on each MMFE board) and the ART data transmission is under Scalable Low-Voltage Signaling for 400 mV (SLVS-400) standard [12]. The transmission speed of each ART channel is 320 Mb/s. A 160 MHz ART Clock, which is synchronized to the global 40 MHz clock, will be provided to the VMM chip to generate the ART signal in DDR mode. After the hit-selection in the ART ASIC, the processed data will be transmitted to the GBTx and then send to the VTTx optic module in GBT widebus format [13].

Besides the miniSAS connectors for the ART signal, there will be another miniSAS connector on board to receive the Bunch Crossing Reset (BCR) signals, the reference clocks and other configuration/control signals from the Level-1 Data Driver Card. The FELIX will send out these configuration and control signals to the 2 ART ASICs and GBTx chips via the slow control chip SCA under HDLC protocol [14] [15]. It will also provide the 40.079 MHz reference clocks to the GBTx chips reference clock input pins. Then the GBTx will generate the 40.079 MHz and 160.316 MHz reference clock for its corresponding ART ASIC. The CERN developed radiation-tolerant DC-DC convertor ASIC 'FEAST' will be used on this board to provide power for all the chips on ADDC [16].

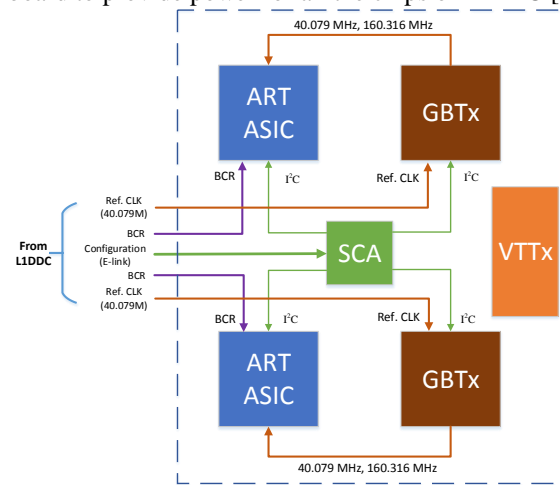

Fig. 3. The block diagram of the ADDC. It has two mirroring paths consist of ART and GBTx chip.

### A. ADDC version 1 prototype (FPGA based)

The 1st version of ADDC prototype was built based ono the Xilinx Artix-7 FPGA and it is at half of the final scale (handles 32 channels of ART input) [17]. At that moment the ART ASIC is still under developing thus the FPGA is used to verify the

ASIC design and excise the communication with the GBTx chip. With the help from the ART ASIC designer group, the ART ASIC core HDL code is migrated to the Artix-7 FPGA to implement the hit-selection algorithm. The VTRx module, which has 1 transmitter for the uplink data transmission and 1 receiver for the down link configuration is used instead of the VTTx module. Through the optical fiber with the QSFP module, the JTAG-based remote configuration to the Artix-7 FPGA is achieved. The photo of the assembled ADDC version 1 prototype is shown in figure 4. There is no dimension restricts on this prototype because it is only used in laboratory. Also the commercial DC-DC convertor LTM4619 was used in this prototype instead of the FEAST module as the latter power module was not released at that moment.

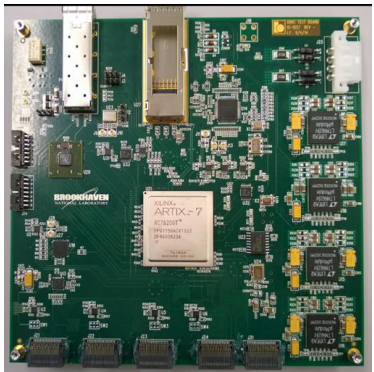

Fig. 4. The photo of the ADDC version 1 prototype. In this version Xilinx Artix-7 FPGA is used to play the role of the ART ASIC. Commercial power chip LTM 4619 is used to replace the FEAST DC-DC module.

### B. ADDC version 2 and version 3 prototypes

In the ADDC version 2 and version 3 prototypes, the dimensions are restricted and all the radiation-tolerant ASICs are evaluated. All the inductors and capacitors are carefully selected to ensure the performance under the radiation hard and strong magnetic field environment. These prototypes are also at the full scale to handle 64 channels of ART inputs. Each ART ASIC has 32 differential input ports to receive the ART signals at 320 Mbps. After hit selection, the result will be transmitted to 14 GBTx e-link ports, which runs under wide-bus mode at 320 Mbps. Then the VTTx module is used to collect the data from the two GBTx chip and transmit them to the trigger processor.

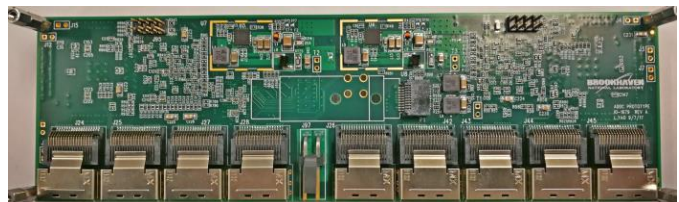
(a)

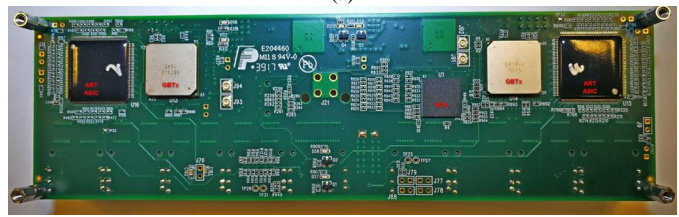
(b)

Fig. 5. (a) The top side of the ADDC the version 3 prototype. (b) The bottom side of the ADDC version 3 prototype. All the custom designed, radiation tolerant ASICs are deployed in this prototype. Cooling plate and foam will be put on the bottom side to make surface contact the chips for heat dissipation

### C. ADDC preproduction prototype

In the version 2 and version 3 prototypes the ART ASIC uses the 144 pin LQFP package. However the yield rate of this package didn't not meet the expectation thus a 128 pin LQFP package is proposed as a replacement. The ADDC preproduction prototype design is revised to match the change of ART ASIC package. Some mounting holes are reserved and makes this board can be also compatible with sockets. With sockets mounted on the board, it can serve as ART chip test board to verify the chip functionality before assembly. Various tests are already planned for the preproduction prototype and once they are finished with positive results the massive production board fabrication will start.

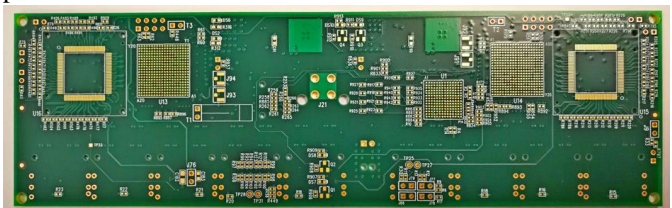

Fig. 6. The photo of the bare PCB for the ADDC preproduction prototype. A 128 pin LQFP is proposed for the ART ASIC to improve its packaging yield rate. This PCB is compatible for the 128 pin chip or the corresponding socket. With the socket it can be used as the chip testing board.

## III. TESTING OF THE ADDC

A FMC based test platform is designed to provide a complete test environment for the ADDC boards. This test platform works with the VC707 commercial develop kit, connects to the ADDC and will provide the simulated ART signals and configuration/clock signals to the ADDC board. With GBT-FPGA compatible firmware running on VC707, it can also receive the two channels of processed data from the VTTx of the ADDC and examine if there is any error during the ART hit-selection or data transmission. The diagram of this test platform and connections to the ADDC is shown in figure 7. There are 8 miniSAS connectors on this FMC test board to provide the simulated ART signals to the ADDC and 1 miniSAS connector to provide the configuration/clock signals. To emulate the configuration communication from FELIX/L1DDC, a GBTx is placed on the FMC test board (emulates the GBTx on L1DDC). This GBTx can be bypassed by sending configuration data directly from the FPGA to another preserved miniSAS (CH 4 in figure 7). However to ensure it simulates the real environment as much as possible this is only a backup plan for early debugging purpose. With the VTRx module, it communicates with the VC707 FPGA that runs a GBT-FPGA compatible firmware (emulates the FELIX). Two SFP module are placed on the FMC board to receive the two optic channels from the VTTx on ADDC. The received data will be transmitted to a computer for data analyze (emulates the trigger processor).

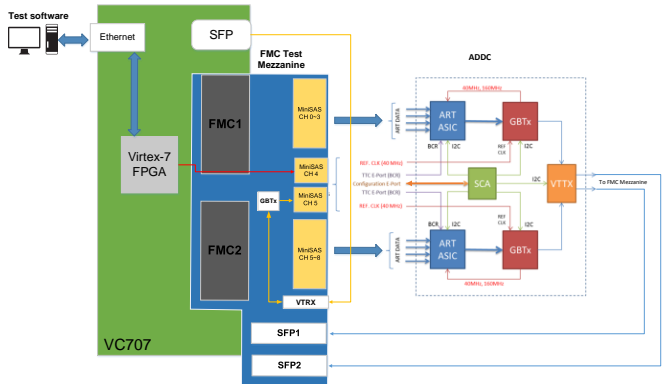

Fig. 7. The block diagram of the ADDC test platform. It can provide complete testing of the ADDC without the involvement of other NSW electronics board.

### A. ADDC test platform firmware design

In the test platform firmware the Microblaze soft processor core will handle all the data transmission to/from the Ethernet port. It handles all the commands and data to/from the computer. Besides that the firmware implements 3 major modules to communicate with the ADDC:

The 1st is the module that distributes the ART signals to the miniSAS ports. The simulated ART data will be generated by the software, it is the parallel data which consists the BCID, ART channel (0~63) and the 6-bit address. This module will translate the simulated ART data to the corresponding miniSAS channel and serial bit stream.

The 2nd one is the configuration module that communicates to the ADDC SCA. Although this module has the links to communicate with the SCA chip directly through the reserved miniSAS connector, in normal tests it will go through the GBTx chip on the FMC mezzanine. The configuration data will be packed in HDLC protocol by the software and then in the firmware it will be organized in the GBT frame.

The 3rd part is a revised version of GBT-FPGA firmware to receive the processed data from ADDC. The DDR module is integrated here to achieve the Ping-Pong buffering structure for the data. Then the data can be read out to the Ethernet under the Direct-Memory-Access function through AXI bus.

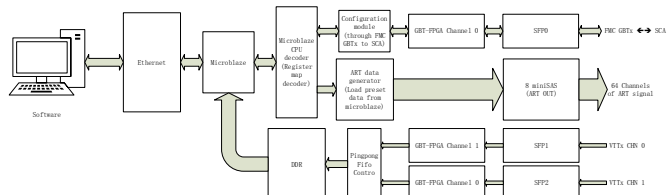

Fig. 8. The block diagram of the test platform firmware. The Microblaze soft processor core is the bridge to communicate with the computer, resolve the commands and send them to the ART signal generator or configuration module while the last function is gathering the buffered data from the DDR module.

### B. ADDC test software design

The software is designed in Python, it provides the configuration to the ADDC and the full test of all ADDC functions. Since the configuration to the ADDC will go through the CERN developed slow control chip SCA, the HDLC protocol is implemented. The ADDC can be configured to be running under several different modes and for each mode the test software will generate corresponding data to send to the ADDC miniSAS inputs and then check the results back from the ADDC fibers.

The flow chart of the test software is shown in figure 9. After started it will first try to build the connection with the VC707 via Ethernet and then configure the GBTx on the FMC mezzanine. Then it will start to configure the GBTx and ART ASICs on the ADDC via the SCA slow control chip. After every power on, the phase relationship between the GBTx/ART might be changed thus the ART ASIC will be configured to send out preset static data to align the output phase. The 64 input channels will also need to be aligned with preset data pattern that can be chosen in the software. After all the chips are configured and the phases aligned, the software will go through each test on the ADDC. If any error is found during these tests, the information will be recorded and the board will be marked.

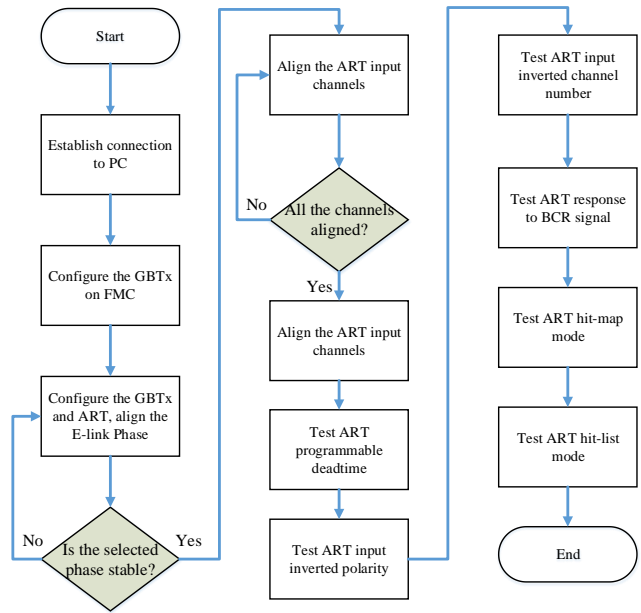

Fig. 9. The flow chart of the test platform software. When the test starts, it will first do the configuration and phase alignment and then step into the test of each aspect. If any error is captured it will be recorded and end the testing.

Figure 10 shows the photo of connections between the FMC test platform and the ADDC board. Currently each ADDC takes about 40 mins to go through all the tests and if necessary this test platform can be used to perform long term test for the ADDC boards to verify the stability.

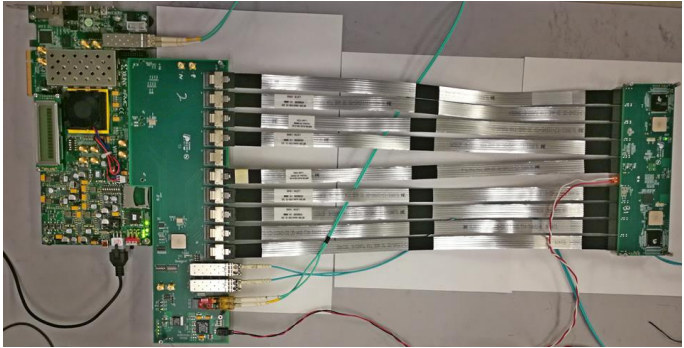

Fig. 10. The photo of the ADDC under testing with the test platform. No other NSW electronics is required when doing the laboratory test of the ADDC.

The ADDC is a key component on the MM trigger path hence the latency performance is a very important factor. Besides the functionality test, the latency of the ADDC has also been measured. The latency of the ADDC can be divided into two parts: the latency of the ART ASIC to process the de-serializing and hit selection and the latency of the GBTx plus VTTx module for the processed data transmission. The scheme of the measurement is described in figure 11.

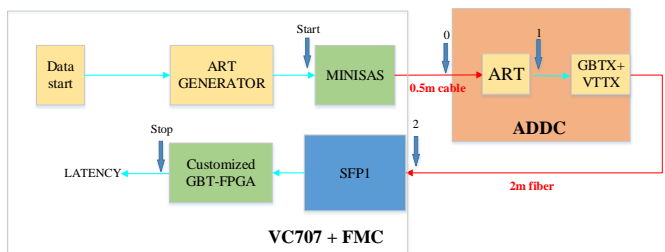

Start to stop: ~254 ns
Start to "1": ~47 ns

Fig. 11. The block diagram of the latency test for the ADDC. Besides the overall latency for the full ADDC board, the detailed latency for the ART ASIC and the GBTx plus VTTx module are also measured.

When the test platform begins to send the serial ART data to the miniSAS connectors connecting with ADDC, it will also generate a flag signal which indicates the start point of the latency. Then after the test platform receives the results from the ADDC, it will also generate a flag signal which indicates the stop point of the latency. To measure the latency for the ART ASIC, we cannot generate this extra signal because its output ports to the GBTx are 320 Mbps serial data in the GBT format. But we can still use the differential probe to directly capture one of its 14 differential output ports on point "1".

The time difference between the start and stop points is measured as around 254 ns and the difference between start and point "1" is around 47 ns. The latency between start and stop point includes not only the ADDC latency but also the propagation delay in the miniSAS connector and 0.5m cable, in the 2m fiber and also the latency of the SFP module and the RX end of the customized GBT-FPGA firmware. The delay in the miniSAS connector and 0.5m cable is estimated to be around 3 ns and the delay in the 2m fiber is estimated to be around 10 ns. The delay of the SFP module and the RX end of the customized GBT-FPGA firmware has been analyzed and measured previously in the test of [18], which is around 54 ns. So after some calculation we have the result that the latency of the ADDC is around 187 ns, which is much less than the required 500 ns. We can also calculate that the latency of the ART ASIC is around 44 ns and the latency of the GBTx plus VTTx module is around 143 ns.

## IV. CONCLUSION

The New Small Wheel deployment in the ATLAS Phase I upgrade is planned to maintain the efficient muon detection with increased event rate. The ART signal, which reflects the 6-bit hit address in the MicroMegas detector, is the trigger primitive. The ADDC board is designed to implement the preliminary hit-selection and transmission of the ART signal to the trigger processor and its performance is a critical part of the trigger system in NSW. The ART chip is a custom designed ASIC to deal with the ART signal on the ADDC. Before this chip is delivered an Artix-7 FPGA has been used on the 1$^{st}$ version of ADDC prototype to evaluate its design and excise the communication with the other radiation-tolerant chips used on ADDC. Since the 2$^{nd}$ version of ADDC prototype all these radiation-tolerant chips, including ART ASIC, have been evaluated and tested. Besides the integration tests of ADDC and other NSW electronics, an FMC based test platform has been designed to provide the individual test environment for the ADDC prototypes and for future mass production ADDC boards. This test platform is controlled by the computer via Ethernet and the Microblaze soft core is used for configuration and data transmission. With the help of this platform all the functionalities of the ADDC have been verified and also the latency performance.